\typein[\sorb]{Enter ``s'' (for small) or ``b'' (for big)}
\input prepictex
\input pictex
\input postpictex
\newlength{\absize}
\if s\sorb \documentstyle{article}
\renewcommand{\baselinestretch}{1.5}
\renewcommand{\arraystretch}{1.5}
\begin{document}
\date{}
\pagestyle{empty}
\thispagestyle{empty}
\renewcommand{\thefootnote}{\fnsymbol{footnote}}
\newcommand{\starttext}{\newpage\normalsize
\pagestyle{plain}
\setlength{\baselineskip}{4ex}\par
\twocolumn\setcounter{footnote}{0}
\renewcommand{\thefootnote}{\arabic{footnote}}
}
\else
\documentstyle[12pt]{article}
\setlength{\absize}{6in}
\setlength{\topmargin}{-.5in}
\setlength{\oddsidemargin}{-.3in}
\setlength{\evensidemargin}{-.3in}
\setlength{\textheight}{9in}
\setlength{\textwidth}{7in}
\renewcommand{\baselinestretch}{1.5}
\renewcommand{\arraystretch}{1.5}
\setlength{\footnotesep}{\baselinestretch\baselineskip}
\begin{document}
\thispagestyle{empty}
\pagestyle{empty}
\renewcommand{\thefootnote}{\fnsymbol{footnote}}
\newcommand{\starttext}{\newpage\normalsize
\pagestyle{plain}
\setlength{\baselineskip}{4ex}\par
\setcounter{footnote}{0}
\renewcommand{\thefootnote}{\arabic{footnote}}
}
\fi
\newcommand{\preprint}[1]{\begin{flushright}
\setlength{\baselineskip}{3ex}#1\end{flushright}}
\renewcommand{\title}[1]{\begin{center}\LARGE
#1\end{center}\par}
\renewcommand{\author}[1]{\vspace{2ex}{\Large\begin{center}
\setlength{\baselineskip}{3ex}#1\par\end{center}}}
\renewcommand{\thanks}[1]{\footnote{#1}}
\renewcommand{\abstract}[1]{\vspace{2ex}\normalsize\begin{center}
\centerline{\bf Abstract}\par\vspace{2ex}\parbox{\absize}{#1
\setlength{\baselineskip}{2.5ex}\par}
\end{center}}
\def\stpltsmbl{\setplotsymbol ({\scriptsize.})}
\newdimen\tdim
\tdim=\unitlength
\renewcommand{\baselinestretch}{1.5}
\renewcommand{\arraystretch}{1.8}
\def\spur#1{\mathord{\not\mathrel{#1}}}
\newcommand{\ol}{\overline}
\newcommand\etal{{\it et al.}}
\newcommand{\ddbar}{D\ol D}
\def\tarrow{\arrow <5\unitlength> [.3,.6]}
\newbox\srd
\setbox\srd=\hbox{\beginpicture
\setcoordinatesystem units <\tdim,\tdim>
\stpltsmbl
\setquadratic
\plot
  0.0   0.0
  4.8  -1.5
  7.5  -5.0
  7.3  -8.5
  5.0 -10.0
  2.7  -8.5
  2.5  -5.0
  5.2  -1.5
 10.0  -0.0
/
\endpicture}
\def\springrd #1 #2 *#3 /{\multiput {\copy\srd}  at
#1 #2 *#3 10 0 /}
\newbox\phdr
\setbox\phdr=\hbox{\beginpicture
\setcoordinatesystem units <\tdim,\tdim>
\stpltsmbl
\setquadratic
\plot
0 0
3 -2.5
0 -5
-3 -7.5
0 -10
/
\endpicture}
\def\photondr #1 #2 *#3 /{\multiput {\copy\phdr}  at
#1 #2 *#3 0 -10 /}

\def\tr{\mathop{\rm tr}\nolimits}

\def\theequation{\thesection.\arabic{equation}}
\preprint{\#HUTP-92/A049\\ 9/92}
\title{$D$-$\ol D$ Mixing in Heavy Quark \\
Effective Field Theory\thanks{Research
supported in part by the National Science Foundation under Grant
\#PHY-8714654.}\thanks{Research supported in part by the Texas National
Research Laboratory Commission, under Grant \#RGFY9206.}
}
\author{
Howard Georgi \\
Lyman Laboratory of Physics \\
Harvard University \\
Cambridge, MA 02138 \\
}
\date{}
\abstract{
I analyze $D$-$\ol D$ mixing using the techniques of heavy quark effect
field theory.~\cite{hqeft} The analysis suggests that the there may be
important cancellations among the dispersive effects of different kinds of
final states, so that the total mixing may be considerably smaller than
previous estimates.~\cite{donoghue,wolfenstein}
}
\starttext

\section{Introduction}

\begin{figure}[htb]
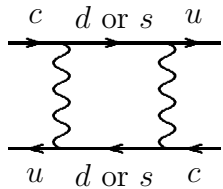

$$
\beginpicture
\setcoordinatesystem units <\unitlength,\unitlength>
\setplotsymbol ({\scriptsize.})
\linethickness=\unitlength
\putrule from -40 20 to 40 20
\putrule from -40 -20 to 40 -20
\photondr -20 20 *3 /
\photondr 20 20 *3 /
\tarrow from -2 20 to 2 20
\tarrow from 2 -20 to -2 -20
\tarrow from -32 20 to -28 20
\tarrow from 28 20 to 32 20
\tarrow from 32 -20 to 28 -20
\tarrow from -28 -20 to -32 -20
\put {$c$} at -30 30
\put {$c$} at 30 -30
\put {$u$} at -30 -30
\put {$u$} at 30 30
\put {$d$ or $s$} at 0 30
\put {$d$ or $s$} at 0 -30
\linethickness=0pt
\putrule from -60 0 to 60 0
\putrule from 0 -40 to 0 40
\endpicture
$$
\caption{\label{boxfigd}The box graph for $D$-$\ol D$ mixing.}
\end{figure}

\begin{figure}[htb]
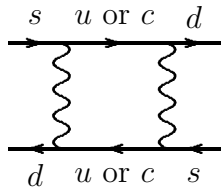

$$
\beginpicture
\setcoordinatesystem units <\unitlength,\unitlength>
\setplotsymbol ({\scriptsize.})
\linethickness=\unitlength
\putrule from -40 20 to 40 20
\putrule from -40 -20 to 40 -20
\photondr -20 20 *3 /
\photondr 20 20 *3 /
\tarrow from -2 20 to 2 20
\tarrow from 2 -20 to -2 -20
\tarrow from -32 20 to -28 20
\tarrow from 28 20 to 32 20
\tarrow from 32 -20 to 28 -20
\tarrow from -28 -20 to -32 -20
\put {$s$} at -30 30
\put {$s$} at 30 -30
\put {$d$} at -30 -30
\put {$d$} at 30 30
\put {$u$ or $c$} at 0 30
\put {$u$ or $c$} at 0 -30
\linethickness=0pt
\putrule from -60 0 to 60 0
\putrule from 0 -40 to 0 40
\endpicture
$$
\caption{\label{boxfigk}The box graph for $K$-$\ol K$ mixing.}
\end{figure}

\bigskip In the classic analyses of $D$-$\ol D$ mixing in the standard model,
Donoghue \etal ~\cite{donoghue} and Wolfenstein~\cite{wolfenstein} argued that
the short distance contributions to the mixing are negligible compared to
``dispersive'' contributions from second order weak interactions with mesonic
intermediate states,
\begin{equation}
\sum_I{\bigl\langle D^0\big|H_W\big|I\bigr\rangle
\,\bigl\langle I\big|H_W\big|\ol{D^0}\bigr\rangle
\over m_D^2-m_I^2+i\epsilon}\,.
\label{dispersive}
\end{equation}
In particular, because of the large $SU(3)$ breaking in $D^0$ decays, the
contributions of particular intermediate states to
(\ref{dispersive}) is apparently much larger than the contribution of the box
diagram, shown in figure \ref{boxfigd}. This is quite different from the
situation in $K$-$\ol K$ mixing, where the box diagram, shown in figure
\ref{boxfigk}, is thought to be comparable to the dispersive contributions.
This difference may be very important phenomenologically. If the dispersive
contribution to $D$-$\ol D$ mixing is as large as suggested in
\cite{donoghue,wolfenstein}, the reach of $D$-$\ol D$ mixing as a signal for
new physics beyond the standard model is reduced. In other words, when
$D$-$\ol D$ mixing is finally observed experimentally, it will be important to
have a good estimate of the standard model contribution, so that we will know
whether we are seeing standard model physics or new physics beyond the
standard model.

Here, we reexamine the issue of $D$-$\ol D$ mixing, making use of the heavy
quark effective field theory (HQEFT).~\cite{hqeft} We will find a puzzle. We
will identify an enhancement of a well-defined set of ``long-distance''
contributions. However, the enhancement does not appear to be as large as
suggested by \cite{donoghue,wolfenstein}. If the heavy quark analysis is at
all valid, it suggests that the total dispersive contribution is considerably
smaller than the contributions found in \cite{donoghue,wolfenstein} from
specific final states. This is certainly possible. Contributions from various
different classes of intermediate states can cancel to give a smaller total
mixing. It may be that the HQEFT analysis implies that such a cancellation
will take place.

When we use the heavy quark effective theory to analyze $D$ meson properties,
we are assuming that $m_c$ is much greater than $\Lambda_{QCD}$. Thus it would
not be completely surprising if the results of the heavy quark analysis were
misleading, but at least it is a well defined approximation from which you
can discuss corrections. At any rate, we will assume in the following that we
can treat the $c$ quark as heavy.

\section{Symmetries of the Standard Model}

Before we discuss the HQEFT analysis in detail, I will discuss the flavor
symmetry structure of $D$-$\ol D$ mixing. I want to make a point of this,
because the enhancements that we find for the long-distance contributions will
be intimately connected to their flavor symmetry properties. The symmetry
structure of the standard model is shown diagrammatically in figure
\ref{symmetry}.

{\tdim=1.25\tdim
\begin{figure}
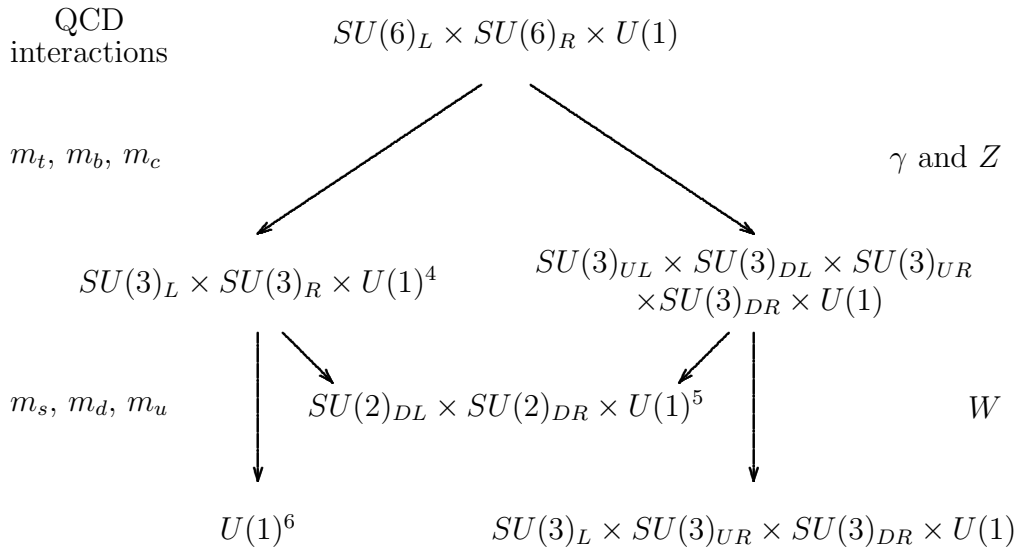

$$
\beginpicture
\setcoordinatesystem units <1.5\tdim,1.5\tdim>
\stpltsmbl
\put {\stack{QCD,interactions}} [l] at -100 100
\put {$SU(6)_L\times SU(6)_R\times U(1)$} at 0 100
\tarrow from -5 90 to -50 60
\tarrow from 5 90 to 50 60
\put {$m_t$, $m_b$, $m_c$} [l] at -100 75
\put {$\gamma$ and $Z$} [r] at 100 75
\put {$SU(3)_L\times SU(3)_R\times U(1)^4$} at -50 50
\put {\stack{$SU(3)_{UL}\times SU(3)_{DL}\times SU(3)_{UR}$,
$\times SU(3)_{DR}\times U(1)$}} at 50 50
\tarrow from -50 40 to -50 10
\tarrow from 50 40 to 50 10
\put {$m_s$, $m_d$, $m_u$} [l] at -100 25
\put {$W$} [r] at 100 25
\put {$U(1)^6$} at -50 0
\put {$SU(3)_{L}\times SU(3)_{UR}
\times SU(3)_{DR}\times U(1)$} at 50 0
\tarrow from -45 40 to -35 30
\tarrow from 45 40 to 35 30
\put {$SU(2)_{DL}\times SU(2)_{DR}\times U(1)^5$} at 0 25
\linethickness=0pt
\putrule from -60 0 to 60 0
\putrule from 0 -10 to 0 110
\endpicture
$$
\caption{\label{symmetry}Quark flavor symmetry.}
\end{figure}}

The $SU(6)_L\times SU(6)_R\times U(1)$ symmetry of the QCD interactions acting
on six massless quarks is broken by the
``large'' quark masses, $m_t$, $m_b$, and $m_c$ down to $SU(3)_L\times
SU(3)_R\times U(1)^4$ and by the interactions of the $\gamma$ and $Z$ down to
$SU(3)_{UL}\times SU(3)_{DL}\times SU(3)_{UR}\times SU(3)_{DR}\times U(1)$.
We will be particularly interested in the common subgroup which is left
invariant by both,
$SU(2)_{DL}\times SU(2)_{DR}\times U(1)^5$, under which the left- and
right-handed $d$ and $s$ transform as doublets. In particular, the
transformation properties of the $\Delta c=1$, charm-changing nonleptonic weak
interactions, $H_W$, under the left- and right-handed $U$-spin,
\begin{equation}
SU(2)_{DL}\times SU(2)_{DR}\,,
\label{uspin}
\end{equation}will play a crucial role in our discussion. The point is that if
we ignore the small mixing to the third family ($s_2,s_3\approx0$), we can
write the $\Delta c=1$ Hamiltonian at the $W$ scale in the form
\begin{equation}
{G_F\over \sqrt2}\;\ol{\psi_L}\gamma^\mu u_L\;\vec\kappa\cdot\vec\tau\;
\ol{c_L}\gamma_\mu\psi_L\,,
\label{hw}
\end{equation}
where $\psi_L$ is the two component $SU(2)_{DL}$ doublet
\begin{equation}
\psi_L=\pmatrix{s_L\cr d_L\cr}
\label{psi}
\end{equation}
and $\vec\kappa$ is the complex constant $U$-spin vector
\begin{equation}
\vec\kappa={1\over2}\pmatrix{c\cr i\cr s\cr}
\label{kappa}
\end{equation}
with
\begin{equation}
c=c_1^2-s_1^2\,,\quad s=2s_1c_1\,.
\label{cabibbo}
\end{equation}

While the detailed form of the weak Hamiltonian changes as you go down to
lower scales,~\cite{gilmanwise} the transformation properties under
(\ref{uspin}) do not, at least not above 1~GeV, because the QCD interactions
do not break the symmetry. Thus the $\Delta c=1$ weak Hamiltonian is
proportional to the $U$-spin vector, $\vec\kappa$, and the $\Delta c=2$ mixing
is proportional to two factors of $\vec\kappa$. This is relevant because
$\vec\kappa$ has the interesting property
\begin{equation}
\vec\kappa\cdot\vec\kappa=0\,.
\label{nilpotent}
\end{equation}
This is the underlying group theoretical statement of the GIM~\cite{gim}
suppression in $D$-$\ol D$ mixing. To make a nonvanishing $U$-spin singlet, we
need a $U$-spin breaking ``spurion'' in the form of an additional $U$-spin
vector $\vec A$ to form a
non-zero combination, $(\vec A\cdot\vec\kappa)(\vec A\cdot\vec\kappa)$. The
obvious $U$-spin vector is related to the $d$-$s$ mass matrix,
\begin{equation}
{\cal M}=\pmatrix{m_s&0\cr0&m_d\cr}\,.
\label{massmatrix}
\end{equation}
However, here is the point. At short distances, the mass matrix,
(\ref{massmatrix}), transforms as a (2,2) under (\ref{uspin}). Thus the
$U$-spin vector must be constructed from the $SU(2)_{DR}$ singlet combination,
${\cal M}{\cal M}^\dagger$, and the $U$-spin violation is
proportional to
\begin{equation}
\tr\left(\tau_3\,{\cal M}{\cal M}^\dagger\right)
=m_s^2-m_d^2\approx m_s^2\,.
\label{uspinshort}
\end{equation}
However, at long distances, the chiral $SU(2)_{DL}\times SU(2)_{DR}$ symmetry
is spontaneously broken by the strong QCD interactions down to the vector
diagonal $U$-spin. At long distances, a $U$-spin triplet can be constructed
from ${\cal M}\Sigma^\dagger$, where $\Sigma$ is the chiral symmetry breaking
condensate. Thus we might expect long distance contributions proportional to
$m_s^3$ and $m_s^2$, rather than the $m_s^4$ that is required for the short
distance contributions.

\section{The Heavy Quark Effective Theory}

The description of $D$-$\ol D$ mixing in the language of the HQEFT that we
will develop in this section may seem peculiar at first. It is well to
remember that the HQEFT is nothing really profound. All the results could be
equally well derived in the conventional QCD theory, so long as we take
careful account of where large momenta are flowing. The HQEFT is simply a
convenience. It allows us to use the framework of effective field theories to
do automatically the work of isolating large momenta in the full theory.
Of course I should admit again that in all of this we are assuming that $m_c$
is large, in particular that there is a gap between $m_c$ and the scale,
$\approx 1$~GeV of chiral symmetry breaking. In reality, there is hardly any
gap at all. Nature has played an exquisite joke on the practitioners of the
HQEFT, because it is really probably only for the $b$ quark that the
techniques of HQEFT apply without very large corrections. Nevertheless, the
HQEFT may be useful even though the corrections are often large, just as
$SU(3)$ and chiral $SU(3)\times SU(3)$ are useful, even though the
$m_s$ is not very small.

When you go below the $c$ mass scale and go to the heavy quark effective
theory, the charm changing nonleptonic operators disappear from the Lagrangian
of the effective theory, because any such term gives rise to a light colored
particle carrying momentum of the order of $m_c$. Thus below the scale $m_c$,
there are no charm changing nonleptonic interactions in the effective theory.
Now that seems a little strange, because the effects of the nonleptonic decays
must still be there. Presumably the way it works is that when you match at
two loops, the heavy quark propagator will acquire a phase because of the
width,
\begin{equation}
{1\over vk-\Delta m-i\Gamma}\,,
\label{width}
\end{equation}
where $\Gamma$ is the width and $\Delta m$ is the difference between the quark
mass and the mass used to define the velocity dependent fields. Thus in this
order, the heavy quark kinetic energy term is complex. It should not
be surprising that the kinetic energy term for the heavy quark in the HQEFT is
non-unitary. Probability is conserved in the HQEFT only if the heavy quark is
absolutely stable. The width, $\Gamma$, describes the leakage of the heavy
quarks out of the HQEFT.

At any rate, the nonleptonic interactions are not there to produce $D$-$\ol D$
mixing in the heavy quark theory. The moral is that there are no
``dispersive'' contributions to the mixing from physics below the scale $m_c$.
In the heavy quark effective theory, all the mixing comes from the matrix
element of the $\Delta c=2$ operators produced by matching at the scale $m_c$.
In the language of the full theory, there is always a momentum of order $m_c$
flowing through the graphs that produce $\ddbar$ mixing. Of course, there are
still two types of contributions:
\begin{enumerate}
\item Contributions from matching of 4-quark operators produced at scales
above $m_c$ (for example, from eliminating the $b$ quark if we do not ignore
$s_2$ and $s_3$);
\item\label{long} Contributions from two insertions of the nonleptonic $\Delta
c=1$ part of the Hamiltonian at the scale $m_c$.
\end{enumerate}
The first are certainly short-distance
contributions. The second must include the long-distance
contributions, but in the heavy quark effective theory in leading order in
$1/m_c$, these effects just contribute to the same kinds of
$\Delta c=2$ operators. As we will see below, the interesting long-distance
contributions are actually non-leading in $1/m_c$.

The type \ref{long} contribution to leading order in $1/m_c$ is a one loop
matching correction calculated by doing the box diagram
calculation with momentum $m_cv$ in the external $c$ lines, and therefore
flowing through the diagram, as shown in figure \ref{boxfigd}. It is easy to
see that this is proportional to (in lowest order in $1/m_c$ and ignoring
$m_d$)
\begin{equation}
{1\over 16\pi^2}\sin^2\theta\cos^2\theta\; G_F^2\;{m_s^4\over m_c^2}
\label{easy}
\end{equation}
so it is small. There is no enhancement. You cannot even use HQEFT to relate
this contribution to $B\ol B$ mixing --- there are two operators, because of
the $v^\mu$ dependence. That is, in addition to the usual operator,
\begin{equation}
\left(\ol{c_v}\gamma^\mu u_L\right)
\left(\ol{\underline{c}_v}\gamma_\mu u_L\right)
\label{usual}
\end{equation}
(where we are using the notation of \cite{tasi} in which
$\underline{c}_v$ represents the heavy antiquark field, there is also a
contribution proportional to the operator
\begin{equation}
\left(\ol{c_v}\spur v u_L\right)
\left(\ol{\underline{c}_v}\spur v u_L\right)
=-\left(\ol{c_v} u_L\right)
\left(\ol{\underline{c}_v} u_L\right)\,.
\label{unusual}
\end{equation}

\begin{figure}[htb]
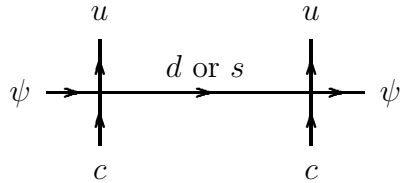

$$
\beginpicture
\setcoordinatesystem units <\unitlength,\unitlength>
\setplotsymbol ({\scriptsize.})
\linethickness=\unitlength
\putrule from -60 0 to 60 0
\putrule from 40 -20 to 40 20
\putrule from -40 -20 to -40 20
\put {$d$ or $s$} at 0 10
\put {$c$} at 40 -30
\put {$c$} at -40 -30
\put {$u$} at 40 30
\put {$u$} at -40 30
\put {$\psi$} at -70 0
\put {$\psi$} at 70 0
\tarrow from 40 -12 to 40 -8
\tarrow from -40 -12 to -40 -8
\tarrow from 40 8 to 40 12
\tarrow from -40 8 to -40 12
\tarrow from -52 0 to -48 0
\tarrow from 48 0 to 52 0
\tarrow from -2 0 to 2 0
\linethickness=0pt
\putrule from -80 0 to 80 0
\putrule from 0 -40 to 0 40
\endpicture
$$
\caption{\label{sixqfig}Feynman graph contributing to tree level matching to
leading order in $\alpha_s$.}
\end{figure}

More interesting are the operators produced by the tree level matching. These
are nonleading in $1/m_c$, but as we will see, they probably give the dominant
contribution to the mixing. First consider mixing to a 6-quark operator from
the diagram shown in figure \ref{sixqfig}.
This gives contributions such as
\begin{equation}
\sin\theta\cos\theta\;{G_F^2\over 2}\; {m_s^2\over m_c^3}\;
\left(\ol{\underline{c}_v}\gamma^\mu \spur v \gamma^\nu u_L\right)\;
\left(\ol{\psi}\gamma_\mu u_L\right)\;\vec\kappa\cdot\vec\tau\;
\left(\ol{c_v}\gamma_\nu \psi_L\right)\,.
\label{6q}
\end{equation}

\begin{figure}[htb]
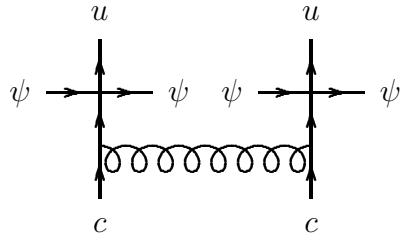

$$
\beginpicture
\setcoordinatesystem units <\unitlength,\unitlength>
\setplotsymbol ({\scriptsize.})
\linethickness=\unitlength
\putrule from -60 0 to -20 0
\putrule from 20 0 to 60 0
\putrule from 40 -40 to 40 20
\putrule from -40 -40 to -40 20
\put {$\psi$} at 10 0
\put {$\psi$} at -10 0
\put {$c$} at 40 -50
\put {$c$} at -40 -50
\put {$u$} at 40 30
\put {$u$} at -40 30
\put {$\psi$} at -70 0
\put {$\psi$} at 70 0
\tarrow from 40 -12 to 40 -8
\tarrow from -40 -12 to -40 -8
\tarrow from 40 -32 to 40 -28
\tarrow from -40 -32 to -40 -28
\tarrow from 40 8 to 40 12
\tarrow from -40 8 to -40 12
\tarrow from -52 0 to -48 0
\tarrow from 48 0 to 52 0
\tarrow from 28 0 to 32 0
\tarrow from -32 0 to -28 0
\linethickness=0pt
\putrule from -80 0 to 80 0
\putrule from 0 -60 to 0 40
%\put {\springru{8}} at -20 -40
%\multiput {\usebox{\srd}} at -40 -20 *7 10 0 /
\springrd -40 -20 *7 /
\endpicture
$$
\caption{\label{eightqfig}Feynman graph contributing to tree level matching to
first order in $\alpha_s$.}
\end{figure}

In addition, there are 8-quark operators produced by graphs like that shown in
figure \ref{eightqfig}. These contributions have no explicit factors of $m_s$.
For example, the contribution of the graph of figure \ref{eightqfig} looks
like
\begin{equation}
{g_s^2G_F^2\over 2m_c^4}\;\left(\ol{\psi}\gamma^\mu u_L\right)\;
\vec\kappa\cdot\vec\tau\;
\left(\ol{\underline{c}_v}\gamma_\alpha \spur v \gamma_\mu T_a u_L\right)\;
\left(\ol{\psi}\gamma_\nu u_L\right)\;\vec\kappa\cdot\vec\tau\;
\left(\ol{c_v}\gamma^\alpha\spur v\gamma^\nu T_a \psi_L\right)\,.
\label{8q}
\end{equation}
Because of the gluon exchange, we will see that contributions of this type are
likely to be less important than the contribution of (\ref{6q}).

As discussed in the previous section, the reason that these contributions can
be important in spite of the extra factors of $1/m_c$ is that the matrix
element of the $U$-spin vector operator $\ol
\psi_L\vec\kappa\cdot\vec\tau\psi_L$
in a low energy hadronic state is proportional to only one power of $m_s$.
This is possible because the chiral $U$-spin symmetry, (\ref{uspin}), is
spontaneously broken. Thus the GIM suppression from these long-distance
contributions is proportional to $m_s\Lambda$ (where $\Lambda\approx 1$~GeV is
the chiral symmetry breaking scale), while the GIM suppression in the leading
operators (including all the short-distance contributions) is guaranteed to be
proportional to $m_s^2$, because the $SU(3)_{DR}$ is unbroken except by the
mass term. Thus while the ``long-distance'' contributions have more factors of
$1/m_c$, they have fewer factors of $m_s$.

In particular, let us use naive dimensional analysis~\cite{powers} (NDA) to
estimate the
matrix elements of the different operators. Dimensional analysis gives the
usual estimate of (\ref{usual}) and (\ref{unusual}) with the NDA relation
\begin{equation}
f_D\approx f\left({\Lambda\over m_c}\right)^{1/2}\,.
\label{f}
\end{equation}
The interesting question is the ratio of the contribution of (\ref{6q}) to
these. The difference is that coefficient of (\ref{6q}) has an extra factor of
\begin{equation}
{16\pi^2\over m_s^2m_c}
\label{extra}
\end{equation}
and the extra U-spin vector operator,
\begin{equation}
d\ol d-s\ol s\,.
\label{operator}
\end{equation}
NDA suggests that (\ref{operator}) has a matrix element of order
\begin{equation}
m_sf^2\,,
\label{operator2}
\end{equation}
so that the ratio of the contribution of (\ref{6q}) to (\ref{usual}) is
expected to be of order
\begin{equation}
{16\pi^2\over m_s^2m_c}m_sf^2={4\pi f\over m_s}{4\pi f\over m_c}\,.
\label{ratio}
\end{equation}
On the other hand, the ratio of the contribution of (\ref{8q}) to
(\ref{usual}) is expected to be proportional to
\begin{equation}
{\alpha_s\over 4\pi}\left({4\pi f\over m_s}{4\pi f\over m_c}\right)^2\,.
\label{ratio8}
\end{equation}

\section{Conclusions}

The relation, (\ref{ratio}), shows a modest enhancement of the
``long-distance'' contributions to $D$-$\ol D$ mixing compared to the
short-distance contributions. Of course, the effect could always be magnified
by an unexpectedly large matrix element of the six quark operator. Certainly,
it is reasonable to expect some additional enhancement, given the large
$SU(3)$ violation in $D^0$ decays. Nevertheless, (\ref{ratio}) and
(\ref{ratio8}) suggest something quite interesting about the calculation of
\cite{donoghue,wolfenstein}. If you look at the contribution of any particular
set of states (related by their $SU(3)$ properties) to the mixing (for
example, all pairs of two light pseudoscalars), {\it ala\/}
\cite{donoghue,wolfenstein}, you would expect a contribution of order $m_s^2$,
because the final states feel the long distance QCD interactions in which the
chiral symmetry is spontaneously broken. But the HQEFT analysis of the
previous section shows that the full order $m_s^2$ contribution to the mixing
is actually suppressed by an additional factor of $\alpha_s/4\pi$. Thus it is
reasonable to expect cancellations between the contributions of different
types of intermediate states. The dominant remaining contribution,
proportional to $m_s^3$, is in some sense intermediate between the short
distance contributions and the ``dispersive'' contributions of
\cite{donoghue,wolfenstein}.

\section*{Acknowledgements}
I am grateful to Chris Carone, Rowan Hamilton, Thorsten Ohl, Gulia Ricciardi
and Elizabeth Simmons for useful discussions. Research supported in part by
the National Science Foundation under Grant \#PHY-8714654 and by the Texas
National Research Laboratory Commission, under Grant \#RGFY9206.

\end{document}